\pdfoutput=1

\newcommand{\flux}      {ergs~cm$^{-2}$~s$^{-1}$~\AA$^{-1}$}
\newcommand{\kms}       {km~s$^{-1}$}
\def\lesssim{\mathrel{\hbox{\rlap{\hbox{\lower4pt\hbox{$\sim$}}}\hbox{$<$}}}}
\def\gtrsim{\mathrel{\hbox{\rlap{\hbox{\lower4pt\hbox{$\sim$}}}\hbox{$>$}}}}
\let\la=\lesssim

\newcommand{\lya}       {Ly$\alpha$}

\documentclass[
    ,final            % use final for the camera ready runs
  ]
  {aipproc}

\layoutstyle{6x9}

\begin{document}

\title[O~VI Absorption in the Galactic Disk, and Future
Prospects]{O~VI Absorption in the Milky Way Disk, and Future Prospects for
  Studying Absorption at the Galaxy-IGM Interface}

\classification{95.85.Mt, 98.35.Hj, 98.38.Kx, 98.58.-w, 98.62.Ra}
\keywords      {Galaxy:disk --- ultraviolet:ISM --- quasars: absorption lines}

\author{D. V. Bowen}{
  address={Dept.~of Astrophysical Sciences, Peyton Hall, Princeton
    University, Princeton NJ~08544}
}

\author{E. B. Jenkins}{
  address={Dept.~of Astrophysical Sciences, Peyton Hall, Princeton
    University, Princeton NJ~08544}
}

\author{T. M. Tripp}{
  address={Dept.~of Astronomy, University of Massachusetts, Amherst, MA~01003}
}

\author{D. G.~York}{
  address={Dept.\ of Astronomy and Astrophysics, University of
  Chicago,  Enrico Fermi Institute, Chicago, IL~60637.}
}

%\author{<author3>}{
%  address={<common address for author2 and author3>}
%  ,altaddress={<author1 address>} % additional visiting address
%}

\begin{abstract}
  We present a brief summary of results from our FUSE program designed
  to study O~VI absorption in the disk of the Milky Way. As a full
  analysis of our data has now been published, we focus on the
  improvements that FUSE afforded us compared to Copernicus data
  published thirty years ago. We discuss FUSE's limitations in
  studying O~VI absorption from nearby galaxies using background QSOs,
  but present FUSE spectra of two probes which indicate the absence of
  O~VI (but the presence of Ly$\beta$) absorption 8 and 63~kpc from a
  foreground galaxy. Finally, we discuss the need for a more sensitive
  UV spectrograph to map out the physical conditions of baryons around
  galaxies.
\end{abstract}

\maketitle

%%%%%%%%%%%%%%%%%%%%%%%%%%%%%%%%%%%%%%%%%%%%
%% MAINMATTER
%%%%%%%%%%%%%%%%%%%%%%%%%%%%%%%%%%%%%%%%%%%%

\section{The FUSE Survey of O~VI Absorption in the Milky Way Disk}

The FUSE survey of O~VI absorption lines in the disk of the Milky Way was a
program that used spectra of 153 early-type stars at latitudes $<
10^{\rm{o}}$ and distances of more than 1~kpc to characterize O~VI absorption
in the plane of the Galaxy. The results from the survey have now been
published in full \citep{bowen08}, so in this contribution, we
highlight just a few of the results from that paper. We begin,
however, in celebrating the accomplishments of FUSE, by comparing some
of the data obtained with the satellite to the data available
prior to its launch.

\subsection{A Copernicus/FUSE comparison}

The results of our FUSE survey were published thirty years and three
months after the seminal survey of \citet{ebj78a} (which built upon
the initial work of \citet{ebj74} and \citet{york74}) using the
Copernicus satellite. Comparing data from the two telescopes might
simply be considered amusing if the goal was to merely demonstrate the
obvious superiority of current instrumentation and detectors over
those available three decades ago. A more serious intent, however, for
such a comparison is to verify the integrity of the older
data. So, for example,
the physical parameters of O~VI absorbing clouds along any particular
sightline should be the same whether measured with Copernicus or
with FUSE. Unfortunately, few lines of sight were actually observed by both
satellites; Copernicus regularly recorded spectra of stars brighter
than $\sim 7$ mag, but these objects were too bright to be observed
with FUSE \citep{sahnow02}.

\begin{figure}[t]
%\hspace*{-1.5cm}\includegraphics[height=.6\textheight]{both_stars}
\hspace*{-1.5cm}\resizebox{1.3\textwidth}{!}
{\includegraphics{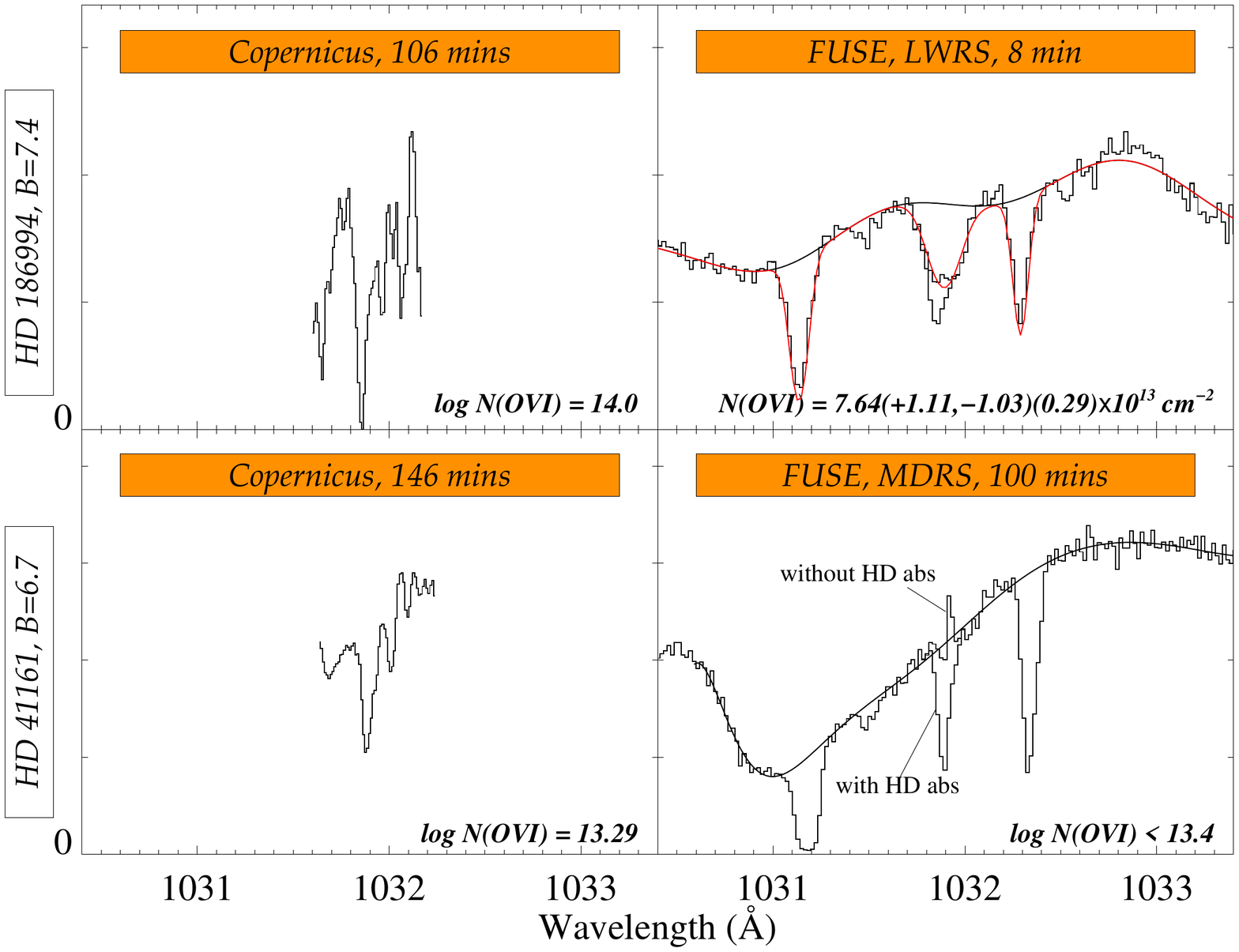}}
  \caption{\label{fig_compare} Comparison of Copernicus
    and FUSE spectra of two stars, HD~186994 (top
    panels) and HD~41161 (bottom panels). 
    The left-hand panels show the coadded Copernicus scans,
    spanning only a small wavelength range; the right hand panels
    show the FUSE spectra, with and without the removal of the HD
    6$-$0 R(0) line from the O~VI~$\lambda 1032$ absorption, 
    as well as the adopted fit to the continuum and,
    for HD~186994, the theoretical Voigt profile fits (red line) to
    the data. The O~VI column densities derived by \citet{ebj78a} for
    Copernicus, and \citet{bowen08} for FUSE, are shown bottom right
    of each panel. For HD~186994, errors in $N$(OVI) are given first
    from continuum fitting, then from counting statistics (see
    \citet{bowen08} for more details). The flux scales are arbitrary,
    and are scaled to allow comparison of the spectra. The strong
    absorption lines flanking the O~VI line are the H$_2$ 6$-$0 P(3) and
    6$-$0 R(4) transitions at 1031.2 and 1032.4~\AA .}
\end{figure}

Nevertheless, in our survey, two stars --- HD~186994 and HD~41161 ---
were observed by both Copernicus and FUSE, and the data from each
satellite are shown in Fig.~\ref{fig_compare}.  This comparison is
somewhat cruel, since with magnitudes of $B=7.4$ and 6.7 for the two
stars, respectively, these were some of the faintest objects
Copernicus could observe. In this respect, the apparently poor
signal-to-noise (S/N) of the two spectra are unrepresentative of the
data used by \citet{ebj78a}. Still, one clear difference
between the two is the much smaller wavelength coverage of the
Copernicus spectra; the satellite used a scanning spectrophotometer to
record stellar spectra while FUSE was fitted with the
multi-spectral-element detector arrays that observers now take for granted.

The coadded Copernicus scans of HD~186994 (taken here from the
Multimission Archive at STScI) represent an exposure time of 106~min;
the 8~min FUSE data taken with the LWRS aperture provide a larger
wavelength range, and so superior a S/N, than the Copernicus spectrum,
that fitting the stellar continuum is much more straightforward.  The
resulting measurement of the physical parameters of the O~VI
absorption [column density $N$(O~VI), Doppler parameter and absorption
velocity] are more precise than those that were derived with
Copernicus, and with the FUSE spectrum we could estimate the
errors in the physical parameters arising from both counting
statistics and errors in the continuum fit. For HD~186994, $\log
N$(O~VI) derived from the Copernicus data was 14.0 or $< 13.85$, as
measured from O~VI~$\lambda 1032$ line or from the O~VI~$\lambda 1037$
line, respectively. These numbers, taken together, are broadly consistent with the
value derived from FUSE, $\log N$(O~VI)$=13.88$, which is reassuring
given the low quality of this particular Copernicus spectrum.

The FUSE data also allow an accurate subtraction of the
HD 6$-$0 R(0) line which contaminates the O~VI profile. This
contamination was well understood by \citet{ebj78a}, and accounted for
in the Copernicus spectrum of HD~41161, where some residual O~VI
absorption was detected after subtraction of the HD line.
Interestingly, in our FUSE spectrum (taken, unusually, using the MDRS
aperture) we determined that all of the observed absorption was due to
the HD line. Nevertheless, our upper limit to $N$(O~VI) is still
consistent with the value given by \citet{ebj78a}.

Verifying the integrity of the earlier Copernicus data is far from
academic: since early-type stars at distances of $d < 1$~kpc were too
bright to be observed with FUSE, we needed the Copernicus data to probe
these distances when examining how O~VI absorption varies with path
length in the Galactic disk.

\subsection{Summary of Program Results}

To explore the characteristics of O~VI absorption in the Galactic
disk, we combined our FUSE observations of stars at $d \sim 1-4$~kpc
with several other datasets. We included stars at $d < 1$~kpc observed
by Copernicus, as well as halo stars \cite{zsargo03}, extragalactic
sightlines \citep{wakker03,savage03}, and nearby white dwarfs
\cite{savage06}, all targeted by FUSE. As noted above, these results are
now published \citep{bowen08}, and our measurements
are available at
\url{http://www.astro.princeton.edu/~dvb/o6home.html}.

Our data confirmed that O~VI absorbing clouds are
ubiquitous throughout the Alpha and Beta quadrants of the Galaxy.  The
O~VI volume density $n$ falls off exponentially with height above the
Galactic plane, as had been shown from previous studies
\cite{widmann98,savage03}. With the FUSE data, however, we were able
to measure the mid-plane density to be precisely
$1.3\times10^{-8}$~cm$^{-3}$, with scale heights of 4.6 and 3.2~kpc for
sightlines in the southern and northern Galactic hemispheres,
respectively. However, even though the O~VI density falls off with
height above the plane, the O~VI absorbing material is not smooth, but
clumpy, with a range of cloud sizes. We were also able to settle a
long standing question as to how much O~VI absorption towards a target
star actually comes from hot circumstellar material around the star
itself --- only a small amount of $N$(O~VI) arises in such regions. We
found that $N$(O~VI) correlates with $d$, demonstrating that O~VI
absorbing clouds are truly interstellar, and composed of many
individual, overlapping, components. The dispersion of $N$(O~VI)
with $d$ is large though, and very different from what would be
expected from absorption by an ensemble of identical clouds. The
velocity extent of O~VI lines follow those of lower ionization
lines observed along the same sightlines, showing that hot and cold
gas are coupled. 

There are different ways to interpret our results, and in the
future, our data should provide the observations necessary to test
theoretical predictions of how hot gas is produced in the Galaxy. We
note that concurrent with our investigations, detailed hydrodynamical
simulations of hot gas in the local Galactic disk were being
engineered by \citet{avillez05}. In these models, the ISM contains a
hot, turbulent multi-phase medium churned by shock heated gas from
multiple supernovae (SNae) explosions. Hot gas arises in bubbles
around SNae, which is then sheered through turbulent diffusion,
destroying the bubbles and stretching the hot absorbing gas into
filaments that dissipate with time.  Although these simulations are
unlikely to be the last word in modelling the hot Milky Way ISM, they
do provide a contemporary context in which to interpret our data. For
example, they successfully predict the mid-plane O~VI density that we
measure in our survey.

\section{O~VI Absorption in other galaxies}

Outside of the Galactic disk, FUSE demonstrated the existence of
copious amounts of O~VI
absorption in the Milky Way halo \citep{wakker03,savage03} and in the
star-forming regions of the Magellanic Clouds
\citep{howk02b,hoopes02}.  O~VI was also detected in Galactic High
Velocity Clouds (HVCs) \citep{sembach_hvcO6}, which posed the
interesting question: how far out from a galaxy can O~VI be detected?
The distances to the HVCs are not well constrained, and O~VI absorbing
HVCs may arise in material infalling into or outflowing from the
Galactic disk (from areas of active star formation, for example), or
further away, from accretion of gas from the intergalactic medium
(IGM) into the extended Milky Way halo, or even the Local Group. In
addition, the relationship between all these local O~VI absorbers, and
the population of weak O~VI absorption systems detected towards QSOs
at redshifts of a few tenths \citep{tripp00a, danforth05, tripp08} is
far from clear. The latter systems may contain a large fraction of
baryons, as much as that currently found in stars, cool gas in
galaxies, and X-ray emitting gas in galaxy clusters. Where these O~VI
lines actually come from, however, is unclear.  Although individual
galaxies have been detected at similar redshifts to O~VI absorption
systems (within impact parameters of $\sim 0.2-1.5$~Mpc
\citep{tripp_0953,savage02,sembach04,tripp06,lehner08}), redshift
information for objects in these fields is incomplete, and the
environment of the absorbing gas is hard to establish at redshifts of
$z > 0.2$.

\begin{figure}[t]
  \vspace*{0cm}\hspace*{0cm}\includegraphics[height=.67\textheight]{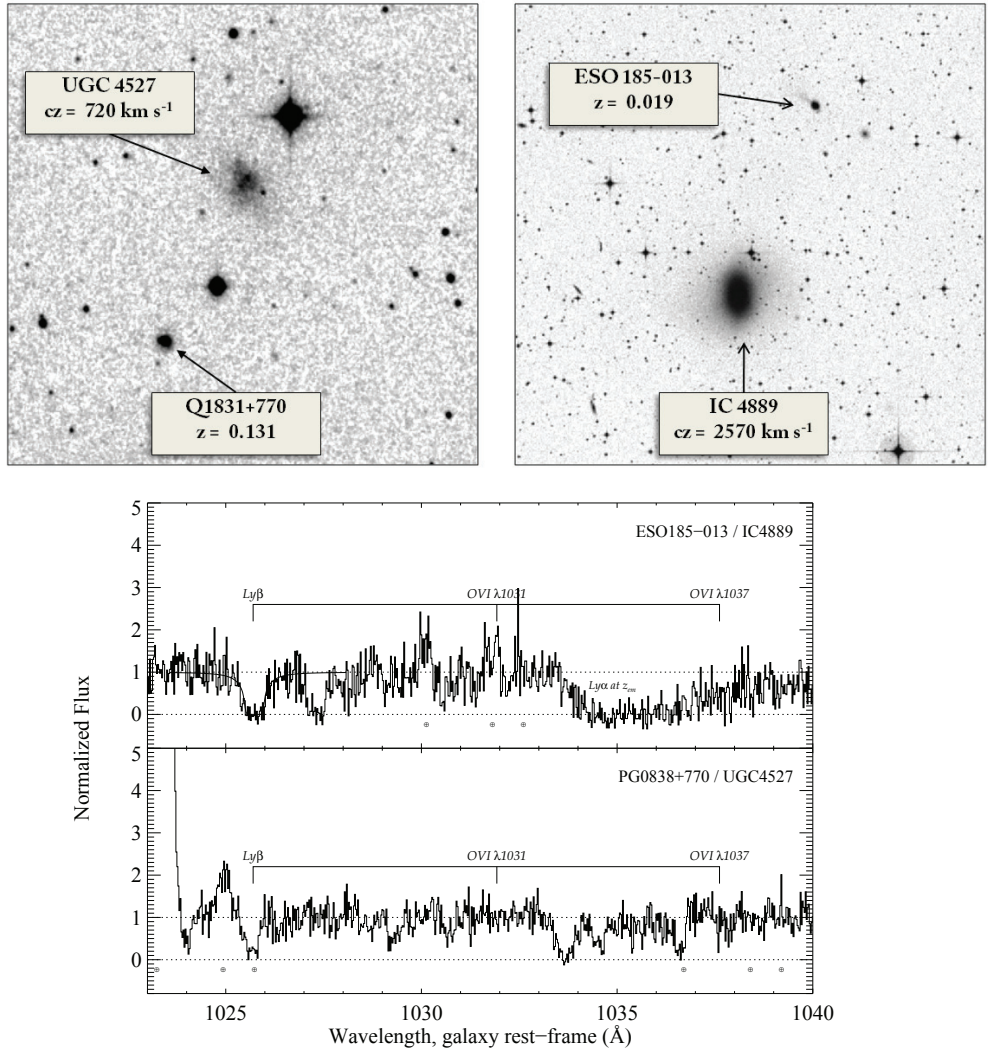}
\caption{\label{fig_fusespec} 
  Spectra of two QSOs that lie close to nearby galaxies, taken as part
  of GI program G020: the top panel shows a 20.0~ksec exposure of
  ESO~185$-$G013 (including data from supplementary program Z909), whose
  sightline passes 63~kpc from IC~4889; the bottom panel shows a
  99.2~ksec exposure of PG~0838+770, whose line of sight passes 8~kpc
  from a low luminosity Im galaxy UGC~4527.  Both spectra are from the
  LiF1A channel, taken using the LWRS aperture, and have been reduced to
  the rest-frame wavelength of the foreground galaxies. The positions of
  Ly$\beta$ and the O~VI doublet are marked, as are the wavelengths
  of strong airglow lines (by $\oplus$ symbols). For ESO~185$-$G013, we
  have drawn a representative theoretical Voigt profile for the Ly$\beta$
  absorption, assuming a Doppler parameter of $b=20$~\kms , and $\log
  N$(H~I)$=19.6$. $N$(HI) could be nearly a dex lower than this
  for higher values of $b$ (see text). The depression at 1035~\AA\ is likely to be
  broad \lya\ absorption at the emission redshift of ESO~185$-$G013.}
\end{figure}

One way to address these questions is to search for O~VI absorption in
the disks and halos of low-$z$ galaxies. Working at low redshifts has
several advantages: there is no ambiguity in the origin of any
detected lines, the properties of the galaxies can be more readily
quantified than at high-$z$, and the physical conditions of the
absorbing gas can be directly linked to those observable properties.
Moreover, the {\it environment} of nearby galaxies -- whether they are
isolated, are interacting with companions, reside in loose groups, or
in clusters --- can be more easily determined than at higher-$z$.

The problem in performing these types of experiments has always been the
difficulty in finding QSOs which are close to galaxies in projection,
{\it and} that are bright enough to be observed in the UV with the available
instrumentation. For FUSE, the challenge was almost insurmountable.
With generous allocations of observing time, of order $100$~ksec, FUSE
could obtain ``adequate'' S/N (at least in the LiF1a channel) of QSOs
with fluxes of $\sim 0.5\times10^{-14}$~\flux . Most of the interesting QSOs close
to low-$z$ foreground galaxies have fluxes less than this, and could
not be targeted by FUSE. Fortunately, there were a few exceptions.

Fig.~\ref{fig_fusespec} shows the results from a program (G020) we
designed to search for Ly$\beta$ and O~VI absorption from the outer
regions of two low-$z$ galaxies. Again, these pairs were selected in
part because the impact parameters between QSO and galaxy were small,
but largely because the QSOs were predicted to have high UV
fluxes. That is, we could not select pairs based on particular galaxy
properties that we might be interested in.
ESO~185$-$013 is an AGN at $z_{\rm{em}}=0.019$ which lies behind the
bright E5 galaxy IC~4889.  The galaxy has a redshift of 2570~\kms, and
the sightline to the AGN passes 63~kpc from its center.  Strong
Ly$\beta$ absorption is detected; unfortunately, the O~VI~$\lambda
1031$ line falls at the position of an O~I$^*$ airglow feature, making
it hard to determine whether the line is present.  Nevertheless,
O~VI~$\lambda 1037$ is not detected to a limit of $\approx 0.15$~\AA .
The H~I column density is difficult to constrain since the Ly$\beta$
line is strongly saturated, and the S/N of the data are not
sufficient to show the onset of damping wings in the line profile.
For Doppler parameters of $b \la 20$~\kms, $\log N$(H~I)$\simeq 19.8$,
but if $b$ is large, e.g., $\sim 30$~kms, $N$(H~I) could be one dex
less. However, the H~I absorption extends in size the structure
seen in 21~cm emission around IC~4889 \citep{oosterloo07} by a factor
of two. The radio data measures $\log N$(H~I) to a limit of $\sim 19$;
our data suggests that $N$(H~I) remains relatively high at a radius
twice that seen at 21~cm.

The second QSO-galaxy pair studied in our program was
PG~0838+770/UGC~4527. The QSO sightline passes only 8~kpc from the
UGC~4527, which is a low surface brightness Im galaxy at a redshift of
720~\kms. We again detect Ly$\beta$ at the redshift of the galaxy,
although it is likely that the line profile is contaminated with
O~I$^{**}$ airglow. However, the non-detection of O~VI is clear, to a
limit of $\approx 0.1$~\AA . Little is known about UGC~4527; deciding
whether the lack of O~VI so close to an irregular dwarf galaxy is
surprising will depend on a better understanding of the galaxy itself,
and ultimately, obtaining data of better quality.

\section{Future Studies of Absorption in galactic disks and halos}

Of course, the study of O~VI provides insights into only one phase of
the gas in and around galaxies. To fully characterize the physical
conditions of gas in galaxy disks and halos, absorption lines from
many different species (each probing gas at different temperatures,
densities, etc.) must be observed. The need for a sophisticated
analysis of what is likely to be a multiphase medium at the boundary
between a galaxy and the IGM is now more compelling than ever, because
our view of galaxies and their relationship to the IGM has changed
dramatically over the last decade.  The exponential growth in
available computing power has allowed detailed modelling of the large scale
structure (LSS) of the Dark Matter (DM) in the universe, along with
its evolution over a significant period of time.
More importantly, these models incorporate the gas hydrodynamics
required to predict the formation and development of galaxies, and
incorporate the likely symbiosis between the physical process at work in the
evolution of a galaxy, and the IGM itself. So, for example, galaxies
must interact and enrich the IGM at all epochs via various feedback
mechanisms: gas may be expelled either from intense bursts of star
formation via strong Galactic winds, or from outflows from a central
AGN. Conversely, the IGM must influence galaxy evolution by the
action of channeling baryons along DM filaments into galaxy groups.
These infusions of gas will most likely change the metallicity of a
galaxy.

Arguably, our ability to test the simulations with observations lags
behind the development of these models. The use of QSO absorption lines
enables us to directly probe the galaxy-IGM interface, but the data
are sparse. There are two obvious goals for future observations.
First, we need to study gas on galaxy scale-lengths around a large
number of galaxies.  For example, we still have little (unambiguous)
information on how the density of gas and its ionization state
declines with radius from a galaxy --- a seemingly fundamental piece
of information for models of galaxy evolution.  Second, we need to
probe individual galaxies along {\it multiple} sightlines, to examine how
the properties we measure for an ensemble of galaxies might actually
vary in a single system. Indeed, mapping the gaseous structures around
single galaxies with multiple lines of sight may be the best way to
determine how gas accretes onto a galaxy and/or how it escapes. 

Studying galaxies with multiple probes can only be accomplished at
low-$z$, where the angular extent of a galaxy is large, and the
background surface density of QSOs is high. On the other hand, probing
the inner regions of nearby galaxies is more difficult, because QSOs
which shine through the hearts of bright low$-z$ galaxies are not
readily detected. Instead, a different approach to exploring the inner regions of
galaxies is to work at a somewhat higher redshift.

\begin{figure}[t]
  \hspace*{0cm}\resizebox{1.0\textwidth}{!}
{\includegraphics{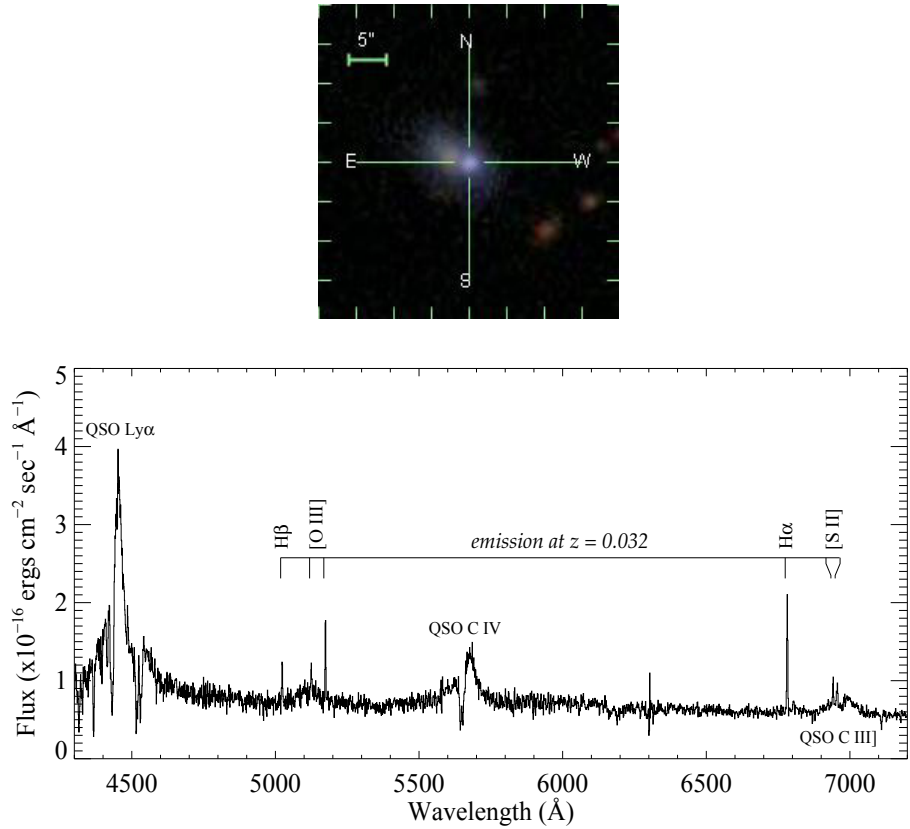}}
\caption{\label{fig_j1042} Identification of a galaxy in front of a
  background QSO using the QSO spectrum. {\it Right:} The flux from
  the $z=2.67$ QSO J104257.58+074850.5 dominates the spectrum taken
  with the SDSS spectrograph --- broad emission lines of \lya\ C~IV
  and CIII] are clearly visible. Superimposed on the spectrum,
  however, are narrow emission lines from a galaxy at $z=0.032$. {\it
    Left:} In
  this case, the intervening galaxy can be seen in SDSS imaging data. 
  Galaxies discovered using this technique are inevitably probed at
  very small impact parameters, and are thus candidates for future studies of the
  inner disks and halos of a wide variety of galaxies. 
}
\end{figure}

Over the last few years we have been identifying galaxies cataloged by
the {\it Sloan Digital Sky Survey} (SDSS)
that lie close to QSOs on the sky, with a emphasis on finding
QSO-galaxy pairs with very small separations. One technique has been to
identify multiple emission lines ([O~II], [O~III], H$\alpha$, etc.)
from low-$z$ galaxies in the spectra of background QSOs. The fibers
used by SDSS to obtain spectra of selected objects are 3$''$ in
diameter, and can collect light from both a QSO and any galaxy along
the line of sight. This ``spectroscopic'' technique has enabled us to
find galaxies at $z\sim 0.2$ probed only a few kpc from their
centers by a QSO (York et~al.~in prep). One of these {\it Galaxies on
  top of QSOs} (GOTOQs) is shown in Fig.~\ref{fig_j1042}.
Studying galaxies close to QSOs at $z\sim 0.2$ is not so easy, compared
to studying QSOs behind $z\sim 0$ galaxies. 
Nevertheless, the GOTOQs offer a special opportunity to
probe gas in the inner regions of galaxies, which can complement the
studies on larger scales discussed above.

The problem in achieving these goals is the same one mentioned in
the previous section --- finding QSOs that are bright enough to be
observed with available satellites. The {\it Cosmic Origins
  Spectrograph} (COS) which will be installed in HST in 2009 will
certainly make significant advances in probing the galaxy-IGM
interface, but difficulties remain. For example, if the goal is to map
low-$z$ galaxies with multiple QSO sightlines, it is quite possible to
find a sufficient number of QSOs beyond several hundred kpc, but at
smaller distances (where much of the galaxy-IGM interaction is
probably taking place) too few QSOs are bright enough for observation
with COS. Further, for studying the inner regions of galaxies, whether
we select nearby galaxies or those at redshifts of a few tenths, the
number of suitable pairs will still be relatively small. Yet in order to
characterize the gas around galaxies selected by their properties --- their
luminosity, morphology, star-formation rates, environment, etc. --- we
will, in the end, need to probe several hundred systems to fully
characterize the galaxy-IGM interface.

Achieving these goals will require a new facility.  The
requirements for the ideal UV spectrograph are obvious, and have been
stated by many previous authors: it would be able to reach
sensitivities of a few $\mu$Jy, about a factor of ten times more
sensitive than COS; it would have a resolution of less than 10~\kms,
to enable accurate measurements of column densities and Doppler
parameters, and permit mapping of the velocity distribution of
multicomponent complexes; and it would cover the entire UV wavelength
range, from 912~\AA\ through to the atmospheric limit of $\sim
3200$~\AA .

With such an instrument, we could map out the physical conditions of
gas on scales ranging from galactic disks to IGM large scale
structures.  Eventually, we would chart the variations in these
conditions over a significant fraction of galactic history, as we
extended our techniques to higher redshifts.  Comparing our results to
simulations which will continue to grow ever more sophisticated
would enable us to understand comprehensively the life-cycle of
baryons in the universe.

%%%%%%%%%%%%%%%%%%%%%%%%%%%%%%%%%%%%%%%%%%%%
%% Sample figure:
%%
%% The option [height=...] scales the picture to the given height,
%% without it it would be printed at its nominal size
%%%%%%%%%%%%%%%%%%%%%%%%%%%%%%%%%%%%%%%%%%%%

%%%%%%%%%%%%%%%%%%%%%%%%%%%%%%%%%%%%%%%%%%%%%%%%
%% BACKMATTER
%%%%%%%%%%%%%%%%%%%%%%%%%%%%%%%%%%%%%%%%%%%%%%%%

\begin{theacknowledgments}
  The work described in this contribution was funded by subcontract
  2440$-$60014 from the Johns Hopkins University under NASA prime
  subcontract NAS5$-$32985, and by LTSA NASA grant NNG05GE26G.
\end{theacknowledgments}



\bibliographystyle{aipproc}   % if natbib is available


\bibliography{bib2,bib_stars}

\IfFileExists{\jobname.bbl}{}
 {\typeout{}
  \typeout{******************************************}
  \typeout{** Please run "bibtex \jobname" to optain}
  \typeout{** the bibliography and then re-run LaTeX}
  \typeout{** twice to fix the references!}
  \typeout{******************************************}
  \typeout{}
 }

\end{document}